\documentclass[parskip=half]{scrartcl}

\title{Running on Raygun}
\author{Alexander Hirsch \and Peter Thoman}
\date{}

\usepackage[margin=2cm,bottom=3cm,footskip=1cm]{geometry}

\usepackage{microtype}
\usepackage{tgtermes}
\addtokomafont{disposition}{\rmfamily}

\usepackage{xcolor}
\definecolor{linkcolor}{cmyk}{1.0, 0.6, 0.0, 0.56}

\usepackage{hyperref}
\hypersetup{
    colorlinks = true,
    citecolor  = linkcolor,
    linkcolor  = linkcolor,
    urlcolor   = linkcolor,
}
\usepackage{graphicx}

\usepackage[numbers]{natbib}
\newcommand{\cpp}{C\texttt{++}}

\begin{document}

\maketitle

\begin{abstract}
    \noindent
    With the introduction of Nvidia RTX hardware, ray tracing is now viable as a general real time rendering technique for complex 3D scenes.
    Leveraging this new technology, we present Raygun, an open source\footnote{\url{https://github.com/W4RH4WK/Raygun} (MIT license)} rendering, simulation, and game engine focusing on simplicity, expandability, and the topic of ray tracing realized through Nvidia's Vulkan ray tracing extension.

    \vspace{1em}
    \noindent
    \includegraphics[width=\linewidth]{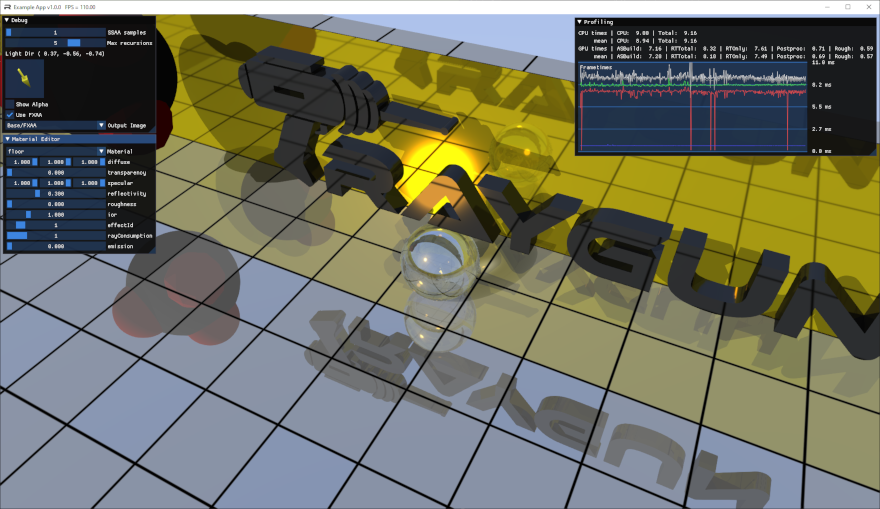}
\end{abstract}

\section{Introduction}

In the beginning, Nvidia RTX~\cite{rtx} could only be used via Nvidia OptiX~\cite{optix}, their dedicated ray tracing application framework, or via Microsoft DXR~\cite{dxr}.
As many developers were hoping for a more open approach to this technology, Nvidia released an (experimental) extension for Vulkan~\cite{vulkan} with their 411.63 graphics driver~\cite{vkray}.
This advancement in technology is what makes Raygun possible.

At the time consumers got their hands on Nvidia RTX capable graphics cards, none of the modern, publicly accessible video game engines allowed using Nvidia RTX for the \emph{whole} rendering process.
Initially we looked at Unity\footnote{\url{https://unity.com}}, Unreal\footnote{\url{https://www.unrealengine.com}}, and Godot\footnote{\url{https://godotengine.org}} just to find out that RTX is not yet available in these engines.
Unity's engine is closed source, Unreal is far too large for us to add RTX support in a tangible amount of time, and Godot has not made the transition (from OpenGL) to Vulkan yet.
This spawned the incipient idea of Raygun.
While Quake II RTX~\cite{quake2rtx}, an implementation of RTX ray tracing using the Quake II engine, already existed at this time, we decided to create our own engine using \cpp{} and modern libraries, focusing on a slim core with expandability in mind.

Nvidia RTX ray tracing works in two steps.
First, a two-layered data structure is instantiated with the vertex information (and transformation matrices) of a scene's 3D models.
Next, a dispatch call initiates the ray tracing process causing the execution of a \textit{ray generation shader}.
This shader controls the initial casting of rays.
Different shaders are invoked depending on whether a ray intersects with any triangle (\textit{closest hit shader} and \textit{any hit shader}) or not (\textit{miss shader}).
This is done for each ray.
The invoked shader has the option to spawn additional rays, continuing the ray tracing process\footnote{The number of these recursive steps is limited by hardware parameters and ray payload size.}.
Each ray can carry a payload (limited, user-defined data) between shaders.

\section{Architecture}

Raygun's main goal is to combine a Vulkan-based ray tracer with other systems typically found in video game engines (e.g.\ scene graph, physics engine, audio system).
It is designed to be used as malleable framework for creating simple, ray traced 3D video games.
The division between game code and engine code can be maintained, as suggested by the file structure\footnote{See subfolders \texttt{raygun} vs.\ \texttt{example}.}.

Abstractions over systems are kept to a minimum so that access to a system's more exotic features is retained.
For instance, we expose the underlying objects of the physics engine so that the physics engine's full potential can be leveraged without adjusting multiple layers of abstractions.
This benefits not only rapid prototyping, but also enables one to swap out systems with little effort at the start of a new project (e.g.\ replacing Nvidia PhysX\footnote{\url{https://developer.nvidia.com/physx-sdk}} with Bullet\footnote{\url{https://github.com/bulletphysics/bullet3}}).

As Vulkan is cross-platform, Raygun targets Windows as well as Linux.
Third-party libraries have been selected accordingly.
CMake\footnote{\url{https://cmake.org}} is used as build system.

\subsection{Render System}

In addition to the commonly used C API, the Khronos Group also provides a \cpp{} API~\cite{vulkanhpp}.
Although using the \cpp{} API can introduce breaking changes across Vulkan SDK versions, it is still preferred to the C API due to Vulkan being a low-level API.
Rendering a simple cube in Vulkan requires a moderate amount of code (compared to high-level APIs like DirectX or OpenGL)~\cite{vulkan_tutorial}.
Code becomes more manageable with the \cpp{} API as language features like RAII\footnote{\url{https://en.wikipedia.org/wiki/Resource_acquisition_is_initialization}}, generic programming, and exceptions are utilised.
While implementing the render system, we introduced thin wrappers around commonly used Vulkan objects (e.g.\ buffers, descriptor sets) to improve the developer experience even more.

The ray tracer is based on the official Nvidia Vulkan ray tracing tutorial~\cite{vkray_tutorial}.
The code has been rewritten using Vulkan's \cpp{} API and now supports rendering multiple objects.
To highlight the benefits of ray tracing over rasterization we extended the shader code to simulate reflection and refraction in addition to shadow casting.

But Vulkan is not just a 3D graphics API, it also provides support for compute workloads.
Considering this, a compute system has been integrated into the engine alongside the render system.
Putting this compute system to use, we also added screen-space roughness approximation and fast approximate anti-aliasing (FXAA).

Shaders are implemented using OpenGL shading language (GLSL) and compiled to SPIR-V during built-time.
Additionally, shaders can be tweaked, recompiled, and reloaded during runtime.

A profiler has been fitted into the engine, using Dear ImGui\footnote{\url{https://github.com/ocornut/imgui}} for visualisation.
Material properties, used by shaders, can be manipulated in real time via an on-screen material editor.

Due to the complexity of Vulkan's API, introducing bugs is quite likely.
To contain this trouble, the Vulkan SDK features a \textit{validation layer} that intercepts and checks Vulkan API calls at runtime.
This mechanism is enabled in Raygun debug builds.
Note that, Vulkan also allows one to tag Vulkan objects with a custom string so they can be identified with ease in profilers and logging output.

\subsection{Other Systems}

Nvidia PhysX is used as physics engine because of its industry roots, license (BSD 3), and visual debugger support.
We added a simple mechanism to register trigger and collision call-backs for physics actors and also forward log messages to the Raygun-wide logging infrastructure\footnote{Using spdlog, see \url{https://github.com/gabime/spdlog}.}

An OpenAL\footnote{\url{https://www.openal.org}} based audio system allows playing of positional sound effects as well as background music.
Audio sources are manipulated through a thin wrapper class as OpenAL uses a very primitive C API.
Sound files are Opus\footnote{\url{https://opus-codec.org}} encoded and stored using the Ogg\footnote{\url{https://www.xiph.org/ogg}} container format.

The aforementioned systems are connected via a basic \textit{scene graph}.
This scene graph is to be used as tree of \textit{entities} (i.e.\ game objects) where each entity holds instance specific data for all systems (e.g.\ model and materials used for rendering).
Strictly speaking, Raygun does not use an entity component system (ECS)\footnote{\url{https://en.wikipedia.org/wiki/Entity_component_system}} architecture, we just adopted the terms \textit{entity} and \textit{system}.

For in-game user interfaces we came up with a UI generator that loosely follows ImGui's API.
Predefined widgets (e.g.\ labels, buttons, sliders) are programmatically composed and automatically arranged.
All UI widgets are ray traced 3D objects --- no textures needed.

Window management and input processing is handled through GLFW\footnote{\url{https://www.glfw.org}}.
Mouse and gamepads are supported, despite using only keyboard input in the provided example application.

3D models are loaded from COLLADA\footnote{\url{https://en.wikipedia.org/wiki/COLLADA}} files using custom JSON documents for material properties.
To minimize material data duplication, materials can inherit properties from other materials.

A dedicated resource manager allows caching of loaded resources.

\section{Future Work}

Raygun is considered feature complete and enables one to experience ray tracing in a developer friendly environment.
Some parts of the code-base may receive additional clean-up and code-quality improvements in the near future.
We intend to keep the engine up-to-date with respect to the Vulkan SDK, leveraging new features given their applicability.

The Khronos Group is working on a vendor independent ray tracing extension for Vulkan.~\cite{vkray_no_nvidia}
As this addition will be based on the currently used, experimental Nvidia extension, migration is expected to be straightforward.
Using a vendor independent extension empowers Raygun to also run on non-Nvidia platforms.

\cpp{}20 will bring lots of interesting features to the language, giving us additional opportunities to further improve code-quality.
Especially \textit{modules}, \textit{designated initializers}, and \textit{ranges} promise to greatly enhance code composition.
\textit{Coroutines}, although stack-less, could facilitate the creation of a scripting system using \cpp{} as embedded domain-specific language (eDSL).

\bibliography{refs}

\end{document}